\documentstyle[aps,prb,twocolumn,floats,epsf]{revtex}

\input{psfig}
\begin{document}
\title{Magnetic exchange interaction induced by a Josephson current}

\author{Xavier Waintal and Piet W. Brouwer}
\address
{Laboratory of Atomic and Solid State Physics, Cornell University,
Ithaca NY 14853, USA \\ {\rm (\today)}
\medskip ~~\\ \parbox{14cm}{\rm	
We show that a Josephson current flowing through a 
ferromagnet--normal-metal--ferromagnet trilayer connected to two
superconducting electrodes induces an equilibrium exchange interaction
between the magnetic moments of the ferromagnetic layers. The sign
and magnitude of the interaction can be controlled by the phase
difference between the order parameters of the two superconductors. We
present a general framework to calculate the Josephson current induced
magnetic exchange interaction in terms of the scattering matrices of the
different layers.  The effect should be observable as the periodic
switching of the relative orientation of the magnetic moments of the
ferromagnetic layers in the ac Josephson effect.
}}

\maketitle


\section{Introduction}

Despite their apparent simplicity,
ferromagnet--normal-metal--ferromagnet 
trilayers exhibit many interesting properties. One example is the
equilibrium
exchange interaction between the magnetic moments of the two
ferromagnets,
which is mediated by the coherent electron motion in the normal
metal spacer layer.\cite{bruno1}
Depending on the thickness of the spacer layer,
this interaction may favor parallel or antiparallel alignment of
the magnetic moments,\cite{bruno1,stiles1} or, in some cases, 
even perpendicular alignment.\cite{slonc1,demokritov} 
In addition, a
torque may be exerted on the magnetic moments when an electrical current
is passed through the trilayer. For this nonequilibrium magnetic 
torque, the preferred magnetic configuration (parallel or 
antiparallel) was found to depend on the sign of the 
current,\cite{slonc3,myers,katine,wmbr} so that a reversal of the current
switches the magnetic moment of the ferromagnets from parallel to
anti-parallel. This reversal can be observed by a measurement of the
conductance, which is larger in the parallel configuration than in 
the antiparallel one (this is known as giant magneto
resistance\cite{ibm}).

A unified description of equilibrium and nonequilibrium torques
can be given using the concept of spin current.
In the case of a ferromagnet--normal-metal--ferromagnet (FNF)
trilayer, this description was introduced by
Slonczewski as an alternative way to calculate the equilibrium
exchange interaction.\cite{slonc2} (The ``standard'' way to
calculate the exchange interaction involves a
computation of the derivative of the free energy to the angle 
between the two magnetic moments.) When electrons scatter from a 
spin-dependent potential, as is appropriate for a mean-field 
description of ferromagnetism, the spin current carried
by the conduction electrons need not be conserved. 
Since the total spin of the system (i.e., the combined
moment of the conduction electrons and the other electrons responsible
for the magnetic moment in the ferromagnetic layers) is conserved,
the lost spin current must have been transferred to the magnetic 
moment of the ferromagnet, which means that a torque is exerted on 
the moments of the
ferromagnets. In this way, the equilibrium
exchange interaction is seen to follow from an equilibrium spin
current flowing from one ferromagnet to the other,\cite{slonc2} 
much like the
equilibrium (persistent) current that exist in mesoscopic metal 
rings,\cite{persistent} 
whereas the nonequilibrium torque arises from the non-conservation
of spin currents flowing in conjunction with the electrical 
current.\cite{slonc3}

The nonequilibrium torque is changed when the FNF trilayer is coupled
to one superconductor (S) contact and one normal-metal (N) contact, 
instead of to two normal-metal contacts. The main difference between
N and S contacts is that the former can carry both spin and charge
currents, while the latter can only carry a charge current for
voltages below the superconducting gap $\Delta_0$. In a previous
publication, we showed that this restriction gives rise to a
nonequilibrium torque that, depending on the direction of the 
electrical current, can lead to the switching of the magnetic
moments to a perpendicular configuration, rather than a parallel
or antiparallel one.\cite{wb} The equilibrium torque (i.e., the
magnetic exchange interaction), however, is
not qualitatively affected by the presence of the one superconducting
contact.\cite{wb2}

In this paper, we consider an FNF junction with two
superconducting contacts.
For this system, a macroscopic supercurrent may flow 
through
the junction already in equilibrium, the magnitude and sign
of the current depending on the phases of the order parameters
of the two superconductors. As ferromagnets break time-reversal
symmetry, they are expected to suppress the Josephson effect.
However for sufficiently thin or
weak ferromagnetic layers
(for instance a Cu$_{1-x}$Ni$_{x}$ alloy~\cite{ryazanov} with $x>0.44$)
, the Josephson effect may survive.
Magnetic Josephson junctions with one ferromagnetic layer
have received considerable attention because of the
possibility of $\pi$-junction behavior,\cite{kulik,shiba,aash,tanaka} 
which has been observed experimentally only recently.\cite{ryazanov,kontos} 
Josephson junctions with
two magnetic layers were studied in 
Refs.\ \onlinecite{bergeret,krivo}, where it was shown that 
the supercurrent for antiparallel alignment of
the magnetic moments can be larger than for parallel alignment.

Here, we consider the exchange interaction in an FNF junction with
two superconducting contacts. We find that the equilibrium exchange 
interaction is 
deeply affected by the presence of the superconductors.
By the same mechanism by which the supercurrent depends on the 
relative orientation
of the two magnetic moments,\cite{bergeret,krivo}
the exchange
interaction depends on the phase difference between the two
superconducting order parameters. 
As a result, the supercurrent controls the exchange interaction 
between the two magnetic moments. 
In contrast to the usual magnetic exchange interaction, which
involves contributions from the whole energy band, the Josephson 
current induced
magnetic exchange interaction is carried only by states
with an energy within a distance of order $\Delta_0$ from the
Fermi energy.
At a first glance, one is tempted to consider the
Josephson current induced torque as the direct analogue of the 
nonequilibrium current-induced torque that exists for normal metal
contacts. However, as we'll show in the remainder of this paper,
that is not the case: Apart from its magnitude, the Josephson
current induced torque has most of the features of the standard
equilibrium magnetic exchange interaction.

This article is devoted to the
study of the Josephson current induced magnetic exchange
interaction and to its
consequences on the dynamics of the magnetic moments.
It is organized as follows:
In Section II, we present the concept of spin current
and discuss the differences
between equilibrium and nonequilibrium torque. We then focus on the 
case of a Josephson junction (both electrodes superconducting). 
Section III contains a brief review of the scattering approach, 
that allows for practical 
calculation of the torques discussed in Section II. We are then ready
to discuss, in Section IV, the magnetic exchange interaction in the
Josephson junction, using various models for the scattering matrices
of the normal and ferromagnetic layers involved. Finally, the effect
of the torque on the dynamics of the magnetic moments is briefly
discussed in Section V. 

\begin{figure}[tbh]
\centerline{\psfig{figure=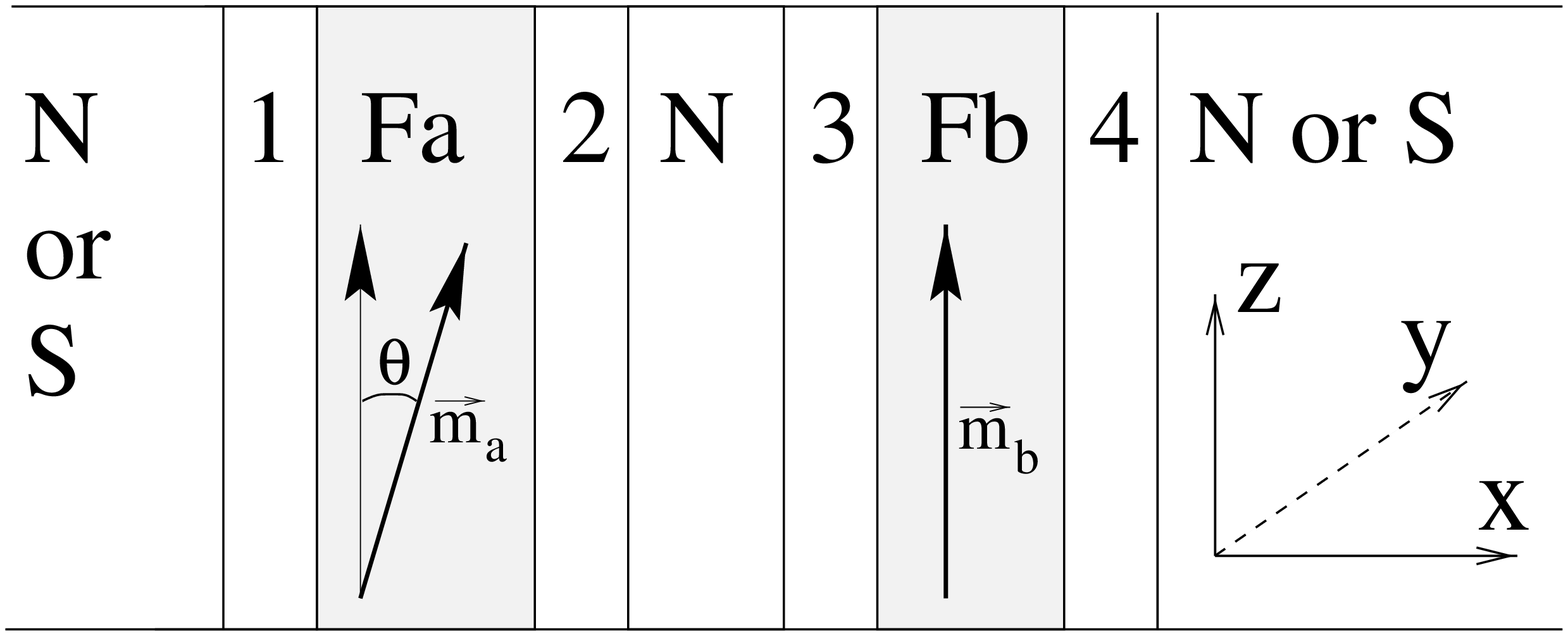,width=7.0cm}}
\vspace{3mm} \narrowtext
\caption[fig2]{
\label{schema}Schematic of the trilayer system considered in this
article. $F_a$ and $F_b$ are the ferromagnetic layers and $\vec{m}_a$ and 
$\vec{m}_b$ their respective magnetic moments. The ferromagnetic
layers  are separated by a normal spacer $N$ and are connected 
to two reservoirs that can be either normal ($N$) 
or superconducting ($S$). The numbers 
$1$, $2$, $3$, $4$  stand for ideal fictitious leads that have been 
added for technical convenience.}
\end{figure}


\section{Spin current and spin torque}

The system we consider is shown in Fig.\ \ref{schema}. It consists
of a ferromagnet--normal-metal--ferromagnet trilayer, connected to 
(possibly superconducting) electrodes on the right and the left. The two
ferromagnetic layers are labeled $F_a$ and $F_b$, the normal metal
spacer layer is labeled $N$. The magnetic
moments of $F_a$ and $F_b$
point in the direction of unit vectors $\hat m_a$ and 
$\hat m_b$, respectively. The angle between $\hat m_a$ and $\hat
m_b$ is $\theta$. We assume that $\hat m_b$ 
points in the $z$-direction. For technical convenience, we have
added pieces of ideal lead (labeled $1$, $2$, $3$, and $4$)
between the layers $F_a$, $N$, and $F_b$, and the reservoirs.

The conduction electrons are described by an effective 
Hamiltonian,\cite{degennes}
\begin{eqnarray}
H_{\rm eff} &=&  \int d\vec{r} \sum_{\alpha,\beta=\uparrow,\downarrow}
\Psi^{\dagger}_{\alpha}(\vec{r}) H_{\alpha \beta} 
\Psi_{\beta}(\vec{r})
\\ && \mbox{}
+ \int d\vec{r} \left[
  \Delta(\vec{r})\Psi^{\dagger}_{\uparrow}(\vec{r})
\Psi^{\dagger}_{\downarrow}(\vec{r})
+ \Delta^*(\vec{r})\Psi_{\downarrow}(\vec{r})\Psi_{\uparrow}(\vec r)
  \right], \nonumber
\end{eqnarray}
where $\Psi^{\dagger}_{\alpha}(\vec{r})$ creates an electron with
spin $\alpha$ and  $\Delta(\vec{r})$ is the superconducting gap. In the
normal regions, $\Delta(\vec{r})=0$. Finally, the $2 \times 2$ matrix
$$
  H= -(\hbar^2/2m) \nabla^2 + V(\vec{r}) - E_F
$$
contains kinetic, potential, and Fermi energy. 
The potential $V(\vec{r})$ represents the spin-independent
scattering from impurities, as well as the spin-dependent
effect of the local exchange field inside the ferromagnets. For
simplicity, we assume that the local exchange field is always parallel
to the total magnetization of the layer (which corresponds to the
neglect of spin-flip scattering). Thus we have
\begin{equation}
V(\vec{r})= e^{-i\frac{\sigma_y \theta(\vec{r})}{2}}
\left(\begin{array}{cc} V_{\rm maj}(\vec{r}) & 0 \\ 
                         0  & V_{\rm min}(\vec{r}) \end{array}\right) 
e^{i\frac{\sigma_y \theta(\vec{r})}{2}}.
\label{eq:Vmajmin}
\end{equation}
Here, maj (min) stands for majority (minority) and $\exp[-i\sigma_y
\theta(\vec{r})/2]$ is a rotation matrix rotating from the external
reference frame to the direction of the local exchange field. Outside
the ferromagnets, $V_{\rm maj} = V_{\rm min}$.
For the system under consideration, the angle $\theta(\vec{r})=\theta$ 
inside $F_a$ and zero elsewhere.  

An expression for the spin current is obtained by writing down the 
conservation equation for the spin density $\vec{\eta}(\vec{r})$,
\begin{equation}
\vec{\eta}(\vec{r})=\frac{\hbar}{2} \sum_{\alpha\beta}
\Psi^{\dagger}_{\alpha}(\vec{r})
\vec{\sigma}_{\alpha\beta} \Psi_{\beta}(\vec{r}),
\end{equation}
$\vec{\sigma}=(\sigma_x,\sigma_y,\sigma_z)^T$ 
being the vector of pauli matrices. The time evolution of spin density
reads,
\begin{eqnarray}
\label{conservation}
\frac{\partial}{\partial t}\langle \vec{\eta}(\vec{r}) \rangle & = &
\frac{i}{\hbar}\langle [H_{\rm eff},\vec{\eta}(\vec{r})] \rangle \nonumber \\
& = &   -\vec{\nabla}\cdot\stackrel{\Rightarrow}{j}(\vec{r}) + 
\frac{i}{2} \langle\Psi^{\dagger}(\vec{r})
[V(\vec{r}),\vec{\sigma}]\Psi (\vec{r}) \rangle,
\end{eqnarray}
with the spin current density tensor $\stackrel{\Rightarrow}{j}$ defined as,
\begin{equation}
\label{spincurrent}
\stackrel{\Rightarrow}{j}(\vec{r}) =  -\frac{i \hbar^2}{4 m} 
\langle \Psi^{\dagger}(\vec{r})\vec{\nabla} \vec{\sigma} \Psi(\vec{r})
 - \vec{\nabla}\Psi^{\dagger}(\vec{r})\vec{\sigma} \Psi(\vec{r})\rangle. 
\end{equation}
The spin current density has one index in spin space and one in real space
while the brackets $\langle\ldots\rangle$ stands for the quantum
mechanical expectation value. Equation (\ref{conservation}) shows
that, unlike charge current, spin current is not conserved inside 
the ferromagnets. In fact, the current induced
torque and the magnetic exchange interaction follow from the spin
current lost by the conduction electrons inside the ferromagnets.
The current induced torque follows from the non-conservaton of
non-equilibrium spin current, while the magnetic exchange interaction
follows from the non-conservation of the equilibrium spin current
between $F_a$ and $F_b$. The total torque $\vec{\tau}_a$, 
$\vec{\tau}_b$ (i.e., the sum of
equilibrium and non-equilibrium contributions) on the layers
$F_a$ and $F_b$ is found as
\begin{equation}
\label{torque}  
\vec{\tau}_a =  \vec{J}_1 - \vec{J}_2 \ \ ,\ \ 
\vec{\tau}_b =  \vec{J}_3 - \vec{J}_4,
\end{equation}
where $\vec{J}_i$ is the total spin current that flows in the $x$ 
direction in region $i$ ($i=1,2,3,4$),
\begin{equation}
\vec{J}_i = \int dy dz \ \vec{j}_x(x,y,z),\ \ \ x \in i.
\end{equation}

\begin{figure}[tbh]
\centerline{\psfig{figure=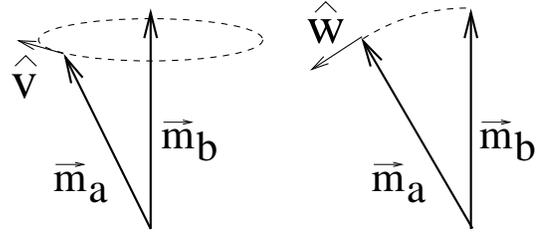,width=7.0cm}}
\vspace{3mm} \narrowtext
\caption[fig2]{
\label{schema2}Schematic of the out of plane component of the
torque $\tau_a^v \hat v$ 
(left side) and the 
in-plane component of the torque $\tau_a^w \hat w$ (right side). 
The equilibrium exchange interaction only has an 
out of plane component while the non equilibrium torque is mainly
in plane.}
\end{figure}

We now focus on the torque $\vec{\tau}_a$ on the magnetic moment of
layer $F_a$. In addition to the unit vector $\hat{m}_a$ that points
along the magnetization direction of $F_a$, we introduce the 
unit vectors $\hat v = \hat m_a \times \hat m_b/|\hat m_a 
\times \hat m_b|$, which points normal to the plane spanned by 
$\hat m_a$ and $\hat m_b$, and $\hat w = \hat m_a \times \hat v$, 
which lies in the plane spanned by $\hat m_a$ and 
$\hat m_b$, but points perpendicular to $\hat m_a$, see Fig.~\ref{schema2}.
For the configuration of Fig.\ \ref{schema}, $\hat v$ is the unit
vector in the $y$ direction and the plane spanned by $\hat m_a$ and 
$\hat m_b$ is the $xz$ plane.
{}From the observation that
\begin{equation}
\left[V(\vec{r}), \sigma_y \right]  = 
-2 i \frac{\partial V}{\partial\theta},
\end{equation} 
combined with Eq.\ (\ref{conservation}), we  find that the
out-of-plane component of the torque $\tau_a^{v} = \vec{\tau}_{a}\cdot
\hat v$ is equal to the derivative of the energy $E=\langle H\rangle$ 
of the trilayer to the angle $\theta$.
No such simple result can be found for the torque in the 
$\hat w$-direction (the component of the torque in the plane spanned
by $\hat m_a$ and $\hat m_b$). Hence, we find for the total 
torque acting on the magnetic moment of $F_a$
\begin{mathletters} \label{torquetot}
\begin{eqnarray}
  \vec \tau_a &=& \tau_a^v \hat v + \tau_a^w \hat w, \\
\label{equivalence}
  \tau_a^v &=& \frac{\partial E}{\partial\theta}, \\
  \tau_{a}^{w} &=& -\case{1}{2}
  \int_{\vec{r}\in F_a}
  d\vec{r}\,
  \langle\Psi^{\dagger}(\vec{r})(V_{\rm maj}-V_{\rm min})
  (\vec \sigma \cdot \hat v) \Psi(\vec{r})\rangle.
\end{eqnarray}
\end{mathletters}%
Since the spin current in the direction of $\hat m_a$ is always
conserved in the absence of spin-flip scattering, 
there is no component of the torque along $\hat m_a$,
cf.\ Eq.\ (\ref{conservation}).
Although the above derivation may seem a little specific, Eq.\
(\ref{torquetot}) can be shown to hold as well in the presence 
of two-body interactions. 

As seen from Eq.\ (\ref{torquetot}), the total torque consists
of two contributions, which are sketched in Fig.\ \ref{schema2}. 
The in-plane torque $\tau^w_a$ directly pushes $\hat m_a$ towards 
or away from $\hat m_b$. It is the
main component of the nonequilibrium torque, which is discussed
in Refs.\ \onlinecite{slonc3,wmbr}. In equilibrium, however, no
spin current can flow outside the trilayer, so that, by Eq.\
(\ref{torque}), 
\begin{equation}
  \vec{\tau}_a^{\rm equ} =-\vec J_2 = -\vec J_3 = -\vec{\tau}_b^{\rm equ}.
  \label{eq:tauequ}
\end{equation}
Combined with the requirement that $\vec \tau_a$ ($\vec \tau_b$) 
is perpendicular to $\hat m_a$ ($\hat m_b$), 
this relation implies that, in equilibrium, 
the in-plane torque $\tau_a^w$
vanishes.\cite{wb2} The out-of-plane torque
$\tau_a^v$ causes a precession of one magnetic moment around the
other one. This is similar to the Larmor precession of the
moments in a (possibly $\theta$-dependent) magnetic field. In
the presence of dissipation, the system will then relax to the
lowest energy configuration, where the energy is minimal and,
hence, by Eq.\ (\ref{equivalence}), the torque zero.

According to
Eqs.\ (\ref{eq:tauequ}) and (\ref{equivalence}), there are two,
equivalent, ways to calculate the equilibrium torque
$\vec \tau_a^{\rm equ}$: As the
equilibrium spin current $\vec J_2$ 
flowing from $F_a$ to $F_b$, or as the 
derivative of the ground state energy $E$ to the angle $\theta$ 
between $\hat m_a$ and $\hat m_b$. (Note that there is a direct
analogy between the equilibrium spin current flowing inside the
trilayer with an angle $\theta$ between the two
magnetic moments and the persistent current in a
mesoscopic ring in presence of an Aharonov-Bohm flux.)
In the remainder of this paper, we concentrate on the equilibrium
torque in the case where both left and right electrodes are superconducting.


The effective Hamiltonian $H_{\rm eff}$ can be diagonalized by
the Bogoliubov transformation, 
 \begin{equation}
\label{bogoliubov}
\Psi_{\alpha}(\vec{r})=\sum_{\epsilon>0} u_{\alpha,\epsilon}(\vec{r}) 
\gamma_{\epsilon} + v^*_{\alpha,\epsilon}(\vec{r}) \gamma^{\dagger}_{\epsilon},
\end{equation}
where the fermion operator $\gamma^{\dagger}_{\epsilon}$ creates an
excitation at energy $\epsilon$ and $u$ ($v$) is the electron (hole)
component of the solution to the Bogoliubov-De Gennes equation
at energy $\varepsilon$,\cite{degennes}
\begin{equation}
\left(\begin{array}{cc} H & i\Delta \sigma_y \\
                  -i\Delta^* \sigma_y & -H^* \end{array}\right)
\left( \begin{array}{c} u \\ v \end{array} \right)
= \epsilon
\left( \begin{array}{c} u \\ v \end{array} \right).
\end{equation}
In principle, 
the superconducting order parameter $\Delta(\vec{r})$ has to be 
calculated self-consistently,
\begin{equation}
\Delta(\vec{r})=-\frac{i}{2} |g(\vec{r})| \sum_{\epsilon>0}
[1- 2 f(\epsilon)] \ v^{\dagger} \sigma_y u,
\end{equation}
where $g(\vec{r})$ is the BCS interaction constant, which is 
finite inside the superconductor and drops abruptly to
zero outside S, and $f(\epsilon)$ is the Fermi function.
Here, we adopt the simple model that 
$\Delta(\vec{r})$ has its bulk value 
$\Delta=\Delta_0 e^{i\phi/2}$ ($\Delta=\Delta_0 e^{-i\phi/2}$) inside 
the left (right) superconducting
reservoirs, while $\Delta=0$ in the normal layers. At the
normal-metal--superconductor interface, 
$\Delta(\vec{r})$ can be approximated by a step function. This
approximation, discussed in Ref.~\onlinecite{likharev}, is valid for
the quasi-one dimensional geometry we consider here.
Up to terms that are independent of $\theta$ and $\phi$, $H_{\rm eff}$
reads
\begin{equation}
H_{\rm eff}=-\frac{1}{2}\sum_{\epsilon>0} \epsilon + 
\sum_{\epsilon>0}\epsilon \ 
\gamma^{\dagger}_{\epsilon}\gamma_{\epsilon}.
\end{equation}
Combining Eqs.\ (\ref{spincurrent}) and (\ref{bogoliubov}), one
obtains for the spin current
\begin{equation}
\label{uvcurrent}
\stackrel{\Rightarrow}{j} = \frac{\hbar^2}{2m} 
{\rm Im}\sum_{\epsilon >0}
\left( u^{\dagger} \vec{\nabla} \vec{\sigma} u-
 v^{T} \vec{\nabla} \vec{\sigma} v^* \right)\langle
\gamma_{\epsilon}^{\dagger}\gamma_{\epsilon}-\case{1}{2} \rangle.
\end{equation}

Equation (\ref{uvcurrent}) was used to calculate the non-equilibrium
torque for an FNF trilayer with one superconducting contact in Ref.\
\onlinecite{wb}. In that case, only eigenstates of the
Bogoliubov-de Gennes equation with energy close to the Fermi level
were involved, so that an analog of the Landauer formula could be derived 
to find the nonequilibrium current-induced torque.
The equilibrium torque, however, is of a very
different nature since it involves all states in the entire conduction
band. This limits theoretical approaches to analytical calculations for
simple model systems~\cite{bruno1} or ab-initio 
numerical simulations.\cite{stiles1} This difficulty
does not occur for the Josephson current induced torque.

In a Josephson junction, the equilibrium current $I$ at finite
temperature is given by the derivative of the free energy $F$ 
of the junction to the
phase difference $\phi$ between the two 
superconductors,\cite{degennes}
\begin{equation}
\label{defI}
I=\frac{2e}{\hbar} \frac{\partial F}{\partial \phi}.
\end{equation}
This equation is very similar to the equation for the equilibrium 
spin current between the two ferromagnetic layers, $J_2 = 
- \tau_a^{\rm equ}$, where Eq.\ (\ref{torquetot}) gives
\begin{equation}
\label{dFdt}
\tau_a^{\rm equ}= \frac{\partial F}{\partial \theta}.
\end{equation}
Combining these last two equations one finds that,
\begin{equation}
\label{current-torque}
\frac{\partial I}{\partial \theta}=\frac{2e}{\hbar}
\frac{\partial\tau_a^{\rm equ}}{\partial \phi}.
\end{equation}
In other words, a $\theta$-dependence of the supercurrent implies
a $\phi$-dependence of the equilibrium torque. This is a very
suggestive result, as it was recently predicted 
that the Josephson current should be very
sensitive to the angle $\theta$ between the two magnetic 
moments.\cite{bergeret,krivo} 
In what follows, we refrain from calculating the
full equilibrium torque but concentrate on it $\phi$-dependent part,
the rationale being that it is precisely the $\phi$-dependent part 
that can be viewed as the supercurrent induced torque. 
As we shall see, the $\phi$-dependent part of the torque only has
contributions from
energies within $\Delta_0$ of the Fermi energy and can, therefore,
be calculated from the scattering properties of the junction at 
and near the Fermi level.

\section{Scattering matrix formalism}

A general review of the scattering matrix formalism can be found 
in Ref.~\onlinecite{beenakker2}, while the particular application
to the calculation of spin currents in FNF trilayers is discussed in 
Ref.~\onlinecite{wmbr}. Below we give a brief compilation of the
necessary formulas for the calculation of the equilibrium torque in
an FNF trilayer with two superconducting contacts.


The trilayer is bounded in the $y$ and $z$ directions, so that the
corresponding degrees of freedom are quantized and give rise to 
$N_{\rm ch}$ propagating modes at the Fermi level, with 
$N_{\rm ch}\sim A/\lambda_F^2$, $A$ being the cross section of the 
junction and $\lambda_F$ the Fermi wave length. Each transverse
mode appears as a left moving mode and as a right moving mode, and with 
components for particle/hole and spin degrees of freedom.
We expand the 
solution of the BdG equation in terms of these modes and describe
wavefunctions in terms of the $4 N_{\rm ch}$-component vectors 
$\Psi_i^{L(R)}$ which are the projection of the wave functions
$(u,v)$ on the left (right) going modes in the ideal lead $i$
($i=1,2,3,4$, see Fig.\ \ref{schema}). 
The layers $F_a$, $F_b$, and $N$ are characterized by 
$8 N_{\rm ch} \times 8 N_{\rm ch}$ unitary scatterings
matrices ${\cal S}_a$, ${\cal S}_b$, and ${\cal S}_{N}$,
respectively,
\begin{eqnarray}
\left(
\begin{array}{c} \Psi_{1}^{L} \\ \Psi_{2}^{R} \end{array} \right)
 &=&  {\cal S}_a
\left(\begin{array}{c} \Psi_{1}^{R} \\ \Psi_{2}^{L}
\end{array}\right), \nonumber \\
\label{defSi}
\left(
\begin{array}{c} \Psi_{2}^{L} \\ \Psi_{3}^{R} \end{array} \right)
 &=&  {\cal S}_N
\left(\begin{array}{c} \Psi_{2}^{R} \\ \Psi_{3}^{L}
\end{array}\right), \\
\left(
\begin{array}{c} \Psi_{3}^{L} \\ \Psi_{4}^{R} \end{array} \right)
 &=&  {\cal S}_b
\left(\begin{array}{c} \Psi_{3}^{R} \\ \Psi_{4}^{L}
\end{array}\right). \nonumber
\end{eqnarray}
Each of the matrices ${\cal S}_i$ ($i=a,b,N$) is further
decomposed into  $4N_{\rm ch}\times 4N_{\rm ch}$ reflection ($r_i$,
$r'_i$) and transmission ($t_i$, $t'_i$) matrices,
\begin{equation}
\label{defSi2}
  {\cal S}_i=\left(\begin{array}{cc} r_i & t'_i \\ t_i & r'_i \end{array}
\right).
\end{equation}
Further, the scattering matrices ${\cal S}_i$ are diagonal in electron-hole 
space,
\begin{equation}
  {\cal S}_i(\epsilon) =  \left(
\begin{array}{cc} {S}_i(\epsilon) & 0 \\
     	            0    & {S}^*_i(-\epsilon) \\
\end{array}
\right), \label{eq:eh}
\end{equation}
whereas, in spin space, 
${\cal S}_N$ is proportional to the $2 \times 2$ identity
matrix, ${\cal S}_b={\rm diag} ( S_{b\uparrow},S_{b\downarrow})$ is
diagonal, and ${\cal S}_a$ reads
\begin{equation}
{\cal S}_a = e^{-i\frac{\sigma_y \theta}{2}}
\left(\begin{array}{cc} S_{a\uparrow} & 0 \\ 
                         0  & S_{a\downarrow} \end{array}\right) 
e^{i\frac{\sigma_y \theta}{2}}.
\end{equation}
Finally, the full scattering matrix ${\cal S}_{\vphantom{FNF}}$ of 
the trilayer can be calculated by
combining ${\cal S}_a$, ${\cal S}_{N}$, and ${\cal S}_b$ using 
Eqs.~(\ref{defSi}) and (\ref{defSi2}). One thus finds
\begin{equation}
\label{defS}
\left(
\begin{array}{c} \Psi_{1}^{L} \\ \Psi_{4}^{R} \end{array} \right)
 = {\cal S}_{\vphantom{FNF}}
\left(\begin{array}{c} \Psi_{1}^{R} \\ \Psi_{4}^{L} \end{array}\right), 
\ \ \ 
{\cal S}_{\vphantom{FNF}} =\left(\begin{array}{cc} r_{\vphantom{FNF}} & t'_{\vphantom{FNF}} \\ 
t_{\vphantom{FNF}} & r'_{\vphantom{FNF}} \end{array}
\right),
\end{equation}
with,
\begin{eqnarray}
t_{\vphantom{FNF}} &=& t_b (1-r'_N r_b)^{-1} t_N \\ && \mbox{} \times \left[
1- r'_a t'_N r_b (1-r'_N r_b)^{-1} t_N -r'_a r_n\right]^{-1} t_a, 
\nonumber \\
r_{\vphantom{FNF}} &=& r_a + t'_a \left[r_N + r_b (1-r'_N r_b)^{-1}\right] t_N
\\ && \mbox{} \times
\left[1- r'_a (r_N + r_b (1-r'_N r_b)^{-1})\right]^{-1} t_a, \nonumber 
\end{eqnarray}
and similar expressions for $r'_{\vphantom{FNF}}$ and $t'_{\vphantom{FNF}}$.

The scattering matrices ${\cal S}_a$, ${\cal S}_{N}$, and ${\cal S}_b$
are the input parameters of our approach. 
While valuable insight can be obtained by using
simple ansatzes for these matrices, detailed knowledge of the
scattering matrices for FN interfaces is available from ab-initio
calculations\cite{stiles2} and analytical results are known
for the statistical distribution of ${\cal S}_{N}$ for the case
of a disordered normal metal spacer.\cite{beenakker2}

At the interfaces with the superconducting contacts, electrons
incoming from the trilayer are reflected as holes, and vice versa.
This process, known as Andreev reflection, is described by the 
$8 N_{\rm ch}$ dimensional scattering matrix ${\cal S}_A$,
\begin{equation}
\label{defSA}
\left(
\begin{array}{c} \Psi_{1}^{R} \\ \Psi_{4}^{L} \end{array} \right)
 = {\cal S}_A
\left(\begin{array}{c} \Psi_{1}^{L} \\ \Psi_{4}^{R} \end{array}\right), 
\ \ \ {\cal S}_A = \left(\begin{array}{cc} r_A({\phi \over 2}) & 0 \\ 
0 & r_A(-{\phi \over 2}) \end{array} \right).
\end{equation}
In the limit that the superconducting gap $\Delta_0$ is much smaller
than the Fermi energy $E_F$, $r_A$ reads, in electron-hole grading,
\begin{equation}
r_A(\phi)=\alpha(\epsilon) \left(\begin{array}{cc} 0 & i\sigma_y e^{i\phi} \\ 
-i\sigma_y e^{-i\phi} & 0 \end{array}\right),
\end{equation}
where $\alpha(\epsilon)=e^{-i \arccos(\epsilon/\Delta_0)}=
\epsilon/\Delta_0 -i\sqrt{1 - \epsilon^2/\Delta_0^2}$ and $\sigma_y$
acts in spin space. The Andreev reflection matrix $r_A$ is diagonal 
in the space of transverse modes.

{}From Eq.~(\ref{defS}) and Eq.~(\ref{defSA}) one finds the equation
\begin{equation}
{\cal S}_A {\cal S} \left( \begin{array}{c} \Psi_{1}^{R} \\ 
\Psi_{4}^{L} \end{array} \right) =
\left( \begin{array}{c} \Psi_{1}^{R} \\ 
\Psi_{4}^{L} \end{array} \right),
\end{equation}
from which it follows that the spectrum of the Josephson junction
is given by the solutions of
\begin{equation}
\label{mouton}
  \det [ 1 - {\cal S}_A(\epsilon) {\cal S}(\epsilon) ] =0.
\end{equation}
Following Ref.~\onlinecite{bb}, one can then express the free energy
 $F$ in terms of ${\cal S}_A$ and ${\cal S}$. Taking derivatives
to the superconducting phase difference $\phi$ and to the angle
$\theta$ one arrives at expressions for the Josephson current $I$ 
and the equilibrium torque $\tau_a^{\rm equ}$, 
\begin{eqnarray}
\label{chevre}
I &=& \frac{2 e}{\hbar} kT \sum_{n=0}^{\infty} {\rm Tr} \frac{
\frac{\partial {\cal S}_A}{\partial\phi}(i\omega_n) {\cal S}(i\omega_n)}
{1 - {\cal S}_A(i\omega_n) {\cal S}(i\omega_n)}, \\
\label{bouc}
\tau^{\rm equ}_a &=&  kT \sum_{n=0}^{\infty} {\rm Tr} \frac{
 {\cal S}_A(i\omega_n) \frac{\partial {\cal S}}{\partial\theta}(i\omega_n)}
{1 - {\cal S}_A(i\omega_n) {\cal S}(i\omega_n)}.
\end{eqnarray}
Here $k$ is the Boltzmann constant, $T$ the temperature, and 
$\omega_n= (2n+1) \pi kT$ are the Matsubara
frequencies, $n=0,1,2,\ldots$. 
In the next section, Eqs.\ (\ref{chevre}) and (\ref{bouc}) are the 
starting
points for our calculations of the supercurrent and magnetic
exchange interaction.

\section{Josephson induced torque}
In this section, we discuss the Josephson current and Josephson-effect
induced torque for various simple models for the
ferromagnetic layers $F_a$ and $F_b$ and the normal spacer layer $N$. 
We consider the case of a short Josephson junction, i.e., we suppose
that the length of the trilayer is smaller than the
superconducting coherence length, or, equivalently, that the inverse
dwell time inside the trilayer (the ``Thouless energy'') is larger
than the superconducting gap $\Delta_0$. For a short junction, we
can neglect the energy dependence of the scattering matrices ${\cal S}_a$,
${\cal S}_b$, and ${\cal S}_N$, and evaluate them at the Fermi level $E_F$.

\subsection{Toy model}
To illustrate the origin of the Josephson effect induced magnetic
exchange interaction, we first describe a simple (toy) model for
the scattering properties of $F_a$, $F_b$, and $N$: We assume that both
majority and minority electrons are transmitted perfectly through the 
two ferromagnetic layer $F_a$ ($F_b$), but pick up phase shifts that
differ by an amount $\beta_a$ ($\beta_b$) as a result of the Zeeman 
coupling to the exchange field inside the magnetic layer. Transmission 
through the normal metal spacer layer is ballistic as well. 
Such an assumption corresponds to a WKB treatment of the
exchange field for the case where the exchange field does not depend
on the transverse direction. The phase shifts depend on the transverse
mode. In terms of the potential $V(\vec r)$ of Eq.\
(\ref{eq:Vmajmin}), they are given by,
\begin{equation}
\beta_{a(b)}=\frac{\sqrt{2m}}{\hbar} \int_{F_{a(b)}} 
 dx \left[ 
\sqrt{E - V_{\rm maj}}-
  \sqrt{E - V_{\rm min}}\right],
\end{equation}
where $E = E_F - E_{\perp}$ is the longitudinal component of the kinetic
energy (which depends on the mode index).
This model is simple enough so that one can solve Eq.~(\ref{mouton})
directly (see, e.g., Ref.~\onlinecite{beenakker3}). Up to terms
independent of $\phi$, the Free energy 
is then given by
\begin{mathletters} \label{ourstot}
\begin{equation}
  \label{ours}
F=
-N_{\rm ch} kT 
  \left\langle \sum_{\pm} \log \cosh 
\frac{ \Delta_0 \cos [(\phi \pm \gamma)/2]}{2 kT}
  \right\rangle_{\rm ch},
\end{equation}
where we abbreviated
\begin{equation}
\label{ours2}
\gamma=\arccos\left(\cos(\beta_a+\beta_b)\cos^2\frac{\theta}{2}
+\cos(\beta_a-\beta_b)\sin^2\frac{\theta}{2}
\right),
\end{equation}
\end{mathletters}%
and where $\langle \ldots \rangle_{\rm ch}$ indicated an average
over the transverse modes. Equation (\ref{ourstot}) 
reduces to the result of Ref.\ \onlinecite{beenakker3} for the case of a
ballistic point contact. 

\begin{figure}[tbh]
\centerline{\psfig{figure=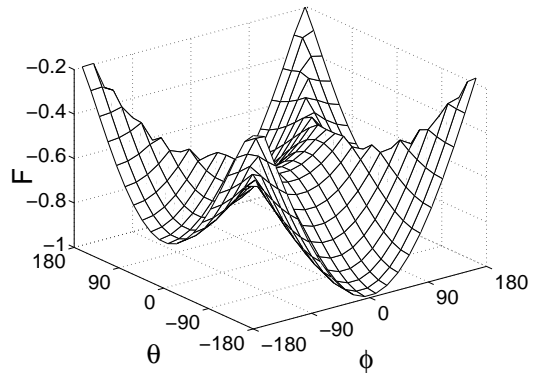,width=7.0cm}}
\vspace{3mm} \narrowtext
\caption[fig2]{
\label{chat} Free energy $F$ as a function of $\theta$ and $\phi$
at zero temperature for the toy model \protect\ref{ourstot}, 
with $\beta_a=\pi/3$ and
$\beta_b=\pi/8$. $F$ is plotted in unit of $N_{\rm ch} \Delta_0$.}
\end{figure}

As an illustration, the contribution
to the zero temperature free energy $F$ from one transverse mode is 
plotted in Fig.~\ref{chat} for a generic choice of the phases 
$\beta_a$ and $\beta_b$. We note that $F$ indeeds depends on both 
$\phi$ and $\theta$. While the $\phi$-dependence of $F$ leads to the 
existence of an equilibrium charge current $I$
through the junction --- the
Josephson current ---, the $\theta$-dependence of $F$ causes an 
equilibrium spin current between $F_a$ and
$F_b$, i.e., a magnetic exchange interaction. We also note that
the $\phi$ and $\theta$ dependencies of $F$ are of comparable
size, which allows us to estimate the equilibrium exchange interaction
as $\tau^{\rm equ} \sim \hbar I_{\rm crit}/2e$, where $I_{\rm crit}$ is the
critical current of the Josephson junction.

The observation
that the $\phi$ and $\theta$-dependences of $F$ are of  
comparable magnitude is valid for arbitrary $\beta_a$ and 
$\beta_b$. However, the
location of the minima and maxima in $F$ depends on whether
$\gamma$ is smaller or larger than $\pi/2$: The minimum of $F$ at
fixed $\theta$ shifts from $\phi=0$ to $\phi=\pi$ if $\beta_a$ and 
$\beta_b$ (and $\theta$) are such 
that $\gamma$ exceeds $\pi/2$. A minimum of $F$ for $\phi=\pi$ 
corresponds to a $\pi$ junction. Since $\gamma$ 
interpolates from $\beta_a+\beta_b$ to $\beta_a-\beta_b$
as one increases the angle $\theta$ between the two magnetic moments,
the $\pi$-junction behavior can be induced by rotating one magnetic
moment
with respect to the other if the phases $\beta_a$ and $\beta_b$ are
sufficiently large.
Similarly, by varying $\phi$, one can switch the minimum of the free 
energy from
$\theta=0$ to $\theta=\pi$, thus favoring a parallel or antiparallel
configuration of the magnetic moments. 

Figure \ref{chat} represents the contribution from only one
transverse channel. As different transverse channels have different
phase shifts $\beta_a$ and $\beta_b$, their contributions to the
supercurrent and to the magnetic exchange interaction do not need
to add up constructively.
In order to
observe an appreciable supercurrent and/or a supercurrent-induced
magnetic exchange interaction, the ferromagnetic layers must be
sufficiently thin or the exchange field must be sufficiently weak
that the phases $\beta_a$ and $\beta_b$ typically do not exceed
unity --- so that all contributions to $I$ or $\tau^{\rm equ}$ 
add up constructively. 
We wish to point out that this
is not an impossible condition to meet. In fact, the same
condition applies to the existence of a supercurrent through a
magnetic Josephson junction with a single ferromagnetic layer. 
Such supercurrents have been observed, see Ref.~\onlinecite{ryazanov} 
Moreover, it
is important to realize that it is only the {\em difference} of
phases between majority and minority electrons that plays a role.
Any common phases are cancelled out as a result of the Andreev
scattering. As a result, the magnitude and sign of the Josephson
current and the exchange interaction do not depend on the phase
shifts picked up in the normal-metal spacer layer.
This is very different from the standard magnetic
exchange interaction, where the sign and magnitude
of the interaction depends
sensitively on the thickness of the normal-metal spacer layer.

\subsection{Chaotic normal layer}

Let us now turn to a more realistic model where reflection processes
occurring inside the ferromagnetic and normal-metal layers and at their
interfaces are fully taken into account, and where scattering from
impurities causes the transverse modes to be mixed. In particular, we 
want to
study the effect of spin filtering: the fact that majority and
minority electrons have different transmission probabilities for
transmission through a ferromagnet layer. 
(Spin filtering is the dominant source of nonequilibrium
current-induced torque for FNF trilayers
with normal-metal contacts.) The Josephson current is
expected to decrease with increasing reflection inside the
FNF trilayer, and with an increasing amount of spin filtering.
(Since the Josephson current is carried by
Cooper pairs, both the minority and the majority electrons must be
transmitted in order to get a current; see Ref.\ \onlinecite{lambert}
for a discussion of this effect in the context of an FNF trilayer
with one superconducting contact.)

Here we assume that the scattering matrix of the normal layer is
drawn from the circular orthogonal ensemble from random matrix
theory, i.e., in particle/hole grading we set 
\begin{equation}
\label{triton}
{\cal S}_{N} = \mbox{diag}\,(S_0 \otimes 
\openone_2,S_0^* \otimes \openone_2),
\end{equation} 
where $\openone_2$ is the 
$2 \times 2$ identity matrix in spin space and
$S_0$ is a $2N_{\rm ch} \times 2N_{\rm ch}$ unitary 
symmetric matrix chosen with 
uniform probability from the manifold of 
$2N_{\rm ch} \times 2N_{\rm ch}$ unitary 
symmetric matrices. The ensemble of scattering matrices corresponds
to an ensemble of FNF trilayers with the same $F_a$ and $F_b$,
but different disorder configurations in $N$. The circular
ensemble is appropriate for a trilayer where
the normal part would be, for example, a dirty metal grain or an
amorphous material.\cite{beenakker2} 
When the transmission probabilities of the
ferromagnetic layers are small, the circular ensemble can be used
for an arbitrary diffusive spacer layer.\cite{beenakker2} Our choice 
of a symmetric 
scattering matrix implies that the amount of magnetic field leaking
into the normal layer from the ferromagnets $F_a$ and $F_b$ must
be sufficiently small, so that time-reversal symmetry is preserved
inside $N$. (If time-reversal symmetry is fully broken in $N$, both
the supercurrent and the Josephson-effect induced torque will
vanish to leading order in $N_{\rm ch}$.)

We are interested in the limit of a large number of channels
$N_{\rm ch}$. (For metals, $N_{\rm ch}$ is already of the order
of $10^3$ for contacts with a width of a few nm.) 
For large $N_{\rm ch}$,
sample-to-sample fluctuations of the current and torque are much
smaller than the ensemble average, so that the ensemble averaged
current $\bar I$ or torque $\bar \tau^{\rm equ}$ is sufficient
to characterize a single sample.
In order to calculate the ensemble average, we first rewrite 
Eq.~(\ref{mouton}) in a slightly different form,
\begin{equation}
\label{mouton2}
{\rm det} [ 1 - {\cal S}_{FS}(\epsilon) S_N ] =0,
\end{equation}
where ${\cal S}_{FS}(\epsilon)$ is the $8 N_{\rm ch} \times 
8 N_{\rm ch}$ matrix describing the combined effect of scattering from 
both ferromagnetic layers backed by the superconductors, as seen from 
the normal spacer layer. The matrix ${\cal S}_{FS}$ has spin, 
particle/hole ($e-h$), channel, and ``$a-b$'' degrees of freedom,
the latter grading referring to whether scattering is from $F_a$ or 
from $F_b$. 
In the $a-b$ grading, ${\cal S}_{FS}(\epsilon)$ reads,
\begin{equation}
{\cal S}_{FS}= \left(\begin{array}{cc} 
  {\cal S}_{{FS},a} & 0 \\ 0 & {\cal S}_{{FS},b} 
  \end{array}
\right),
\end{equation} 
with
\begin{equation}
{\cal S}_{{FS},a} = r'_a + t_a \frac{1}{1-r_A(\phi/2) r_a} r_A(\phi/2) t'_a,
\end{equation} 
and
\begin{equation}
{\cal S}_{{FS},b} = r_b + t'_b \frac{1}{1-r_A(-\phi/2) r_b} r_A(-\phi/2) t_b.
\end{equation}  
With these notations, the ensemble averaged Josephson current reads,
\begin{equation}
\overline{I}= \frac{2 e}{\hbar} kT \sum_{n=0}^{\infty} {\rm Tr}
\ \frac{\partial {\cal S}^{-1}_{FS}}{\partial\phi}({\cal
S}_{FS} - \overline{{\cal G}}),
\end{equation}
with 
\begin{equation}
\label{truite}
\label{TRUITE}
  {\cal G}=  \frac{1}{1- {\cal S}_{FS} {\cal S}_N} {\cal S}_{FS}.
\end{equation}
The average $\overline {\cal G}$ is computed in
the appendix using the method of
Ref.~\onlinecite{brouwer1}. The results of that calculation is a 
self-consistent equation for $\overline{\cal G}$, analogous to the Dyson
equation for the average Green function in a standard impurity 
average,
\begin{mathletters} \label{eq:dyson} \label{EQ:DYSON}
\begin{equation}
\label{chimpanze}
  \overline{{\cal G}} = \frac{1}{ 1 - {\cal S}_{FS} \Sigma}
  {\cal S}_{FS}
\end{equation}
where
\begin{equation}
\label{ouistiti}
\Sigma= {1 \over 2 {\bf P}(\overline{{\cal G}})}\left( 
\sqrt{1 + 4 [{\bf P}(\overline{{\cal G}})]^2} - 1\right)
\end{equation}
and ${\bf P}$ is a projection operator. In $e-h$ space, ${\bf P}$
reads
\begin{eqnarray}
  {\bf P}\left(\begin{array}{cc} A_{ee} & A_{eh} \\ A_{he} & A_{hh} \end{array}
\right) &=&  {1 \over 2 N_{\rm ch}}
 \mbox{tr}_{N_{\rm ch},ab}
\left(\begin{array}{cc} 0  & A_{eh}  \\  A_{he}  & 0 \end{array}
\right) \nonumber \\ && \mbox{} \otimes 1_{ab}\otimes 1_{N_{\rm ch}},
\label{perroquet}
\end{eqnarray}
\end{mathletters}%
where the trace $\mbox{tr}_{N_{\rm ch},ab} \ldots$ is taken in channel 
and $a-b$ space, but not in spin space or particle/hole space.

Equation (\ref{eq:dyson}) reduces to a self-consistent equation for
the $4 \times 4$ matrix $\Sigma$. This equation remains fairly 
complicated and, in general, has to be solved
numerically, even when ${\cal S}_{FS}$ is diagonal in channel space. 
In the limit where ${\cal S}_{FS}$ is close to the 
identity matrix, 
i.e., when both ferromagnetic layers are poorly transparent and
reflect majority and minority electrons with almost the same
reflection phase, a further simplification of Eq.\ (\ref{eq:dyson})
is possible. Assuming
that ${\cal S}_{FS}$ is diagonal in channel space (i.e., the 
ferromagnetic layers do not mix channels), expanding 
${\cal S}_{FS}=1+\delta {\cal S}_{FS} +O(\delta {\cal S}_{FS})^2$, and
defining the $4 \times 4$ matrix
$X=1/(2N_{\rm ch}) \mbox{tr}_{N_{\rm ch},ab} 
\delta {\cal S}_{FS}$,  Eq.~(\ref{eq:dyson}) reduces to
\begin{eqnarray}
X_{he} + \Sigma_{he}X_{ee} +
X_{hh}\Sigma_{he} + \Sigma_{he}X_{eh}\Sigma_{he}  &=& 0,
\nonumber \\
\label{kangourou}
X_{eh} + \Sigma_{eh}X_{hh} +
X_{ee}\Sigma_{eh} + \Sigma_{eh}X_{he}\Sigma_{eh} &=& 0,
\end{eqnarray}
while $\Sigma_{ee}=\Sigma_{hh}=0$
since $\Sigma ={\bf P}(\Sigma)$. Equation (\ref{kangourou}) shows
that for opaque FN interfaces and for diffusive scattering from the
normal spacer layer, there is only a restricted number of parameters
(i.e., the free parameters of $X_{eh}$; $X_{he}$ is related to $X_{eh}$
by particle-hole symmetry)
that determines the supercurrent and the equilibrium torque.

We could only obtain a solution in closed form in the case
where transmission and reflection amplitudes of the ferromagnetic 
layers were real, i.e., still allowing different transmission and
reflection probabilities for majority and minority electrons 
(spin filtering), but without spin-dependent phase shifts in $F_a$ 
and $F_b$. Introducing the mode-averaged transmission probabilities
$T_{a \uparrow}$ ($T_{a \downarrow}$) for majority (minority) 
electrons, 
$$ T_{a \uparrow} = {1 \over N_{\rm ch}} \mbox{tr}\, 
t_{a\uparrow}^{\vphantom{\dagger}} t_{a\uparrow}^{\dagger},\ \
T_{a \downarrow} = {1 \over N_{\rm ch}} \mbox{tr}\, 
t_{a\downarrow}^{\vphantom{\dagger}} t_{a\downarrow}^{\dagger},
$$ 
the spin-averaged
transmission probability $G_{a} = (T_{a \uparrow} + T_{a
\downarrow})/2$, and the geometric mean
$\gamma_{a} = (T_{a \uparrow} T_{a \downarrow})^{1/2}$,
with similar definitions
for $T_{b \uparrow}$, $T_{b \downarrow}$, $G_b$, and $\gamma_{b}$, we find
\begin{eqnarray}
\label{goelan}
\overline{I} &=& \frac{2e}{\hbar} kT N_{\rm ch} \Delta_0 
  \sum_{n=0}^{\infty}
\frac{\sin\phi}{\sqrt{\Delta_0^2+\omega_n^2}} 
  \\ && \mbox{} \times
\frac{\gamma_a \gamma_b }{
\sqrt{\gamma_a^2 + \gamma_b^2 
+ 2 \gamma_a\gamma_b \cos\phi  +
(\omega_n/\Delta_0)^2 (G_a +G_b)^2 }}.
\nonumber
\end{eqnarray}
Equation (\ref{goelan}) is the generalization of Equation (24) of
Ref.~\onlinecite{bb} to the case of contacts with spin-dependent
transmission, in the short junction limit (Thouless energy much 
larger than superconducting gap $\Delta_0$). In the limit of zero
temperature, it reduces to a complete elliptic integral of the first kind.
We note that the Josephson current (\ref{goelan}) does not depend
on the angle $\theta$, so that there is no equilibrium torque in
this case. Since the assumption underlying Eq.\
(\ref{goelan}) was that the electrons do not pick up phase shifts
in the ferromagnetic regions, we conclude that spin-filtering
alone (i.e., the fact that majority and minority electrons have
different transmission probabilities) is not enough to create a
(Josephson current induced) equilibrium magnetic exchange interaction.
We numerically checked that this conclusion still holds when 
${\cal S}_{FS}$ is not close to unity.
 
\subsection{Numerical results}

For a numerical solution of Eq.\ \ref{eq:dyson} that accounts for the fact
that majority and minority electrons experience different phase shifts
while scattering from the ferromagnetic layers, it would be desirable
to have detailed knowledge of the scattering matrices $S_a$ and $S_b$
of the ferromagnetic layers. These scattering matrices can, in
principle, be calculated from ab-initio calculations, see, e.g.,
Ref.\ \onlinecite{stiles2}. 
However,  complete data for all amplitudes (phase shifts and 
probabilities)
are not available in the literature. Therefore, 
 we choose an ansatz for $S_a$ and 
$S_b$ that is close in spirit to the toy model of Sec. IV A. The use
of a simple ansatz can be partly justified by Eq.\ (\ref{kangourou}),
which shows that it is only a finite number of parameters that 
determines the supercurrent and exchange interaction, not the entire
matrix $S_a$ or $S_b$.
For our ansatz, 
we assume that these scattering matrices are diagonal in channel
space and we neglect any channel dependence. We further assume that 
most of the reflection processes take place at the FN interface.
Without loss of generality, we may set the phase picked up by 
minority electrons while traversing the ferromagnetic layers equal to 
zero. Then the difference between minority and majority electrons is
fully described by the phases $\beta_{a(b)}$ picked up by majority 
electrons. This leads us to the ansatz (in left-mover/right-mover
space)
\begin{eqnarray}
 S_{b\uparrow,ee} &=&
\left(\begin{array}{cc} 
 \sqrt{1-T_{b\uparrow}}           & i\sqrt{T_{b\uparrow}}e^{i\beta_b} \\
i\sqrt{T_{b\uparrow}}e^{i\beta_b} &   \sqrt{1-T_{b\uparrow}}e^{2i\beta_b}
\end{array}\right) \otimes \openone_{N_{\rm ch}},
  \nonumber \\ 
 S_{b\downarrow,ee} &=&
  \left(\begin{array}{cc} 
 \sqrt{1-T_{b\downarrow}}           & i\sqrt{T_{b\downarrow}} \\
i\sqrt{T_{b\downarrow}} &   \sqrt{1-T_{b\downarrow}}
\end{array}\right) \otimes \openone_{N_{\rm ch}},
  \label{eq:Sb}
\end{eqnarray}
and  similar equations for $S_{a\uparrow,ee}$ and
$S_{a\downarrow,ee}$, while the scattering matrices
$S_{b\uparrow,hh}$, $S_{b\downarrow,hh}$, 
$S_{a\uparrow,hh}$, $S_{a\downarrow,hh}$
are given by the complex conjugates, cf.\ Eq.\ (\ref{eq:eh}).

\begin{figure}[tbh]
\vspace{-2cm}
\centerline{\psfig{figure=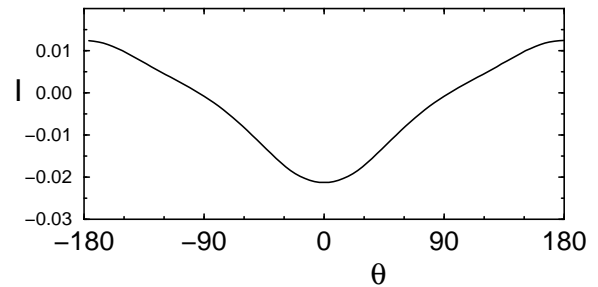,width=8.0cm}}
\vspace{3mm} \narrowtext
\caption[fig2]{\label{fig:3} Supercurrent $I$ at phase
difference $\phi = \pi/2$ between the superconducting
orderparameters, as a function of
the angle $\theta$ between the moments of $F_a$ and $F_b$.
The currents is measured in units of $2e N_{\rm ch}\Delta_0/\hbar$.
We set $T_{a\uparrow}=0.68$, $T_{a\downarrow}=0.29$, $T_{b\uparrow}=0.68$,
$T_{b\downarrow}=0.29$, following ab-initio studies of Ref.\
\protect\onlinecite{stiles2} 
for a Co-Cu-Co trilayer. The phases $\beta_a$ and 
$\beta_b$ were arbitrary set at $\beta_a=35\pi/180$ and
$\beta_b = 115\pi/180$. $kT=0.1 \Delta_0$.}
\end{figure}

With this model for the scattering matrices $S_a$ and $S_b$, 
a typical plot of the supercurrent at $\phi = \pi/2$ as a function
of $\theta$ is shown in Fig.\ \ref{fig:3}. We have taken the
values of the parameters $T_{a\uparrow}$, $T_{a\downarrow}$,
$T_{b\uparrow}$, and $T_{b\downarrow}$ from
realistic estimates for a Co-Cu-Co trilayer,\cite{stiles2,propoverbeux}
while we fixed the phases $\beta_a$ and $\beta_b$ arbitrarily.  
Although the choice
of parameters is specific, the observation that the Josephson
effect induces a magnetic exchange interaction between the
ferromagnetic layers was found to hold for any generic choice
of scattering parameters. [The only exception being the case 
discussed around Eq.\ (\ref{goelan}), for which all scattering 
phase shifts are either $0$ or $\pi$.] Further, we found that when
the phase difference between minority and majority electrons
becomes of order unity, the variation of the supercurrent $I$
with $\theta$ is of the order of the critical current, so that,
up to a numerical factor, the magnitudes of maximum equilibrium
torque and critical current are related as $\tau^{\rm equ}
\sim \hbar I_{\rm crit}/2e$.

\begin{figure}[tbh]
\centerline{\psfig{figure=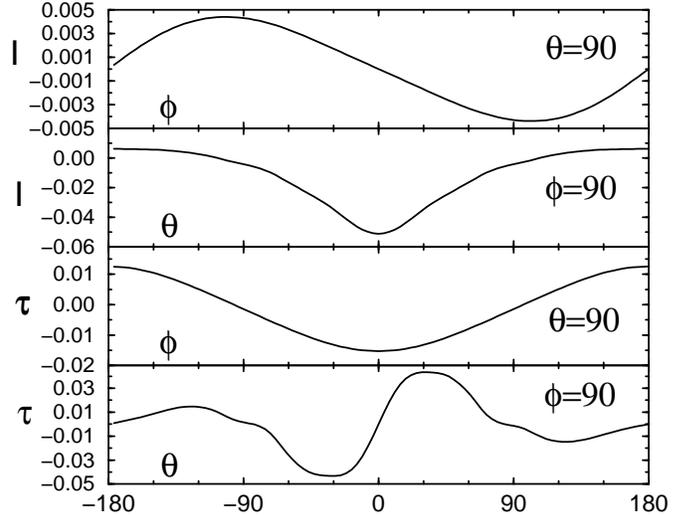,width=9.0cm}}
\vspace{3mm} \narrowtext
\caption[fig2]{
\label{mammouth} Josephson current $I$ and equilibrium torque
$\tau^{\rm equ}$ for the model (\protect\ref{isotropy}) for the
scattering matrix of the normal-metal spacer layer, and with the
same choice of $S_a$ and $S_b$ as in Fig.\ \protect\ref{fig:3}.
From up to down, the panels contain: $I(\phi)$ for $\theta=\pi/2$, 
$I(\theta)$ for $\phi=\pi/2$,
$\tau^{\rm equ}_a(\phi)$  for $\theta=\pi/2$ and
 $\tau^{\rm equ}_a(\theta)$  for $\phi=\pi/2$. For the choice of
parameters used for this figure, the Josephson current
and the equilibrium torque can be fitted with the phenomenological
relation (\protect\ref{tigre}), with
$J_1=0.007$ and $J_2=0.025$. Torques are given in units of 
$\Delta_0 N_{\rm ch}$ and currents in units of 
$2e\Delta_0 N_{\rm ch}/\hbar$. $kT=0.1 \Delta_0$.}
\end{figure}

We now turn to a slightly different model for the normal layer
which we study by doing the disorder average numerically using  
Eq.~(\ref{chevre}).
In this model,~\cite{StoneReview} which
was also used in Refs.\ \onlinecite{wmbr} and \onlinecite{wb}, 
${\cal S}_N$ is given 
by Eq.~(\ref{triton}), where the
$2 N_{\rm ch} \times 2 N_{\rm ch}$ scattering matrix $S_0$ is
parameterized, in a-b grading, as
\begin{equation}
  S_0 =
  \left( \begin{array}{ll} 0 & U \\ U^{\rm T} & 0 \end{array} \right).
  \label{isotropy}
\end{equation}
Here $U$ is an $N_{\rm ch} \times N_{\rm ch}$ unitary matrix,
uniformly distributed in the group of unitary $N_{\rm ch} \times
N_{\rm ch}$ matrices. This model, in which the normal metal
spacer mixes the transverse modes, but does not cause any 
backscattering is appropriate, e.g., for rough FN interfaces. 
As the
same model was considered for quantitative estimates in 
Refs.\ \onlinecite{wmbr} and \onlinecite{wb}, this choice for
${\cal S}_N$ can be used for a quantitative comparison of
the Josephson-effect induced equilbrium torque and the
nonequilibrium torques considered in Refs.\ 
\onlinecite{wmbr} and \onlinecite{wb}. Results are shown in 
Fig.\ \ref{mammouth} for the same choice of parameters as in
Fig.\ \ref{fig:3}. 

For all choices of $S_a$ and $S_b$ that we considered, we find 
that the results are well described
by the phenomenological relation
\begin{equation}
\label{tigre}
\frac{\partial\tau_a^{\rm equ}}{\partial \phi}
=\frac{\hbar}{2e} \frac{\partial I}{\partial \theta}
\approx N_{\rm ch}
\Delta_0  \sin\phi (J_1 \sin\theta+ J_2 \sin2\theta),
\end{equation}
where the constants $J_1$ and $J_2$, which are analogous to the 
quadratic and bi-quadratic coupling constants in the standard 
magnetic exchange interaction, depend on the detailed
choices for $S_a$ and $S_b$. Several properties of this 
phenomenological relation are worth while mentioning.
(i) The torque induced by the Josephson current is proportional
    to the number of transverse channels $N_{\rm ch}$, and hence to 
    the width
    of the trilayer $N_{\rm ch}$. (This property holds if the 
    ferromagnets are sufficiently weak or thin, so that the
    phase difference experienced by majority and minority spins
    is $\lesssim 1$, see the discussion at the end of Sec.\ IV A.) 
    This should be contrasted with the
    regular exchange interaction which does not increase with
    increasing $N_{\rm ch}$ for a disordered normal-metal 
    spacer.\cite{wmbr} Thus, 
    for wide junctions, the Josephson torque
    is parameterically larger than the standard magnetic
    exchange interaction. 
(ii) The torque $\tau^{\rm equ}$ vanishes at $\theta=0$ and 
$\theta=\pi$, irrespective of $\phi$, since for these
angles only spin currents parallel to the magnetic moments
play a role, which are conserved. 
(iii) Similarly, the Josephson current $I$ vanishes at
$\phi=0$ and $\phi=\pi$, irrespective of $\theta$. 
(iv) The Josephson current depends on the angle $\theta$. In the 
case shown in Fig.\ \ref{mammouth}, the junction shows 
$\pi$-junction behavior for $\theta=0$, which disappears when
$\theta$ approaches $\pi$. The relative strength of the
$\theta$-dependent and $\theta$-independent part of the current varies
with the phases $\beta_a$ and $\beta_b$, and the $\pi$-junction 
behavior is not
necessarily there for all choices of $\beta_a$ and $\beta_b$. 

\section{Conclusion and discussion}

We have shown that for a ferromagnet--normal-metal--ferromagnet (FNF)
trilayer coupled to two superconducting contacts, the Josephson
effect enhances and controls the magnetic exchange interaction
between the magnetic moments of the two ferromagnetic layers. 

The Josephson-effect induced torque bears important similarities
and differences with the nonequilibrium and equilibrium torques
in an FNF trilayer with two normal metal contacts:
\begin{itemize}
\item
  The Josephson-effect induced torque points perpendicular to
  the plane spanned
  by the directions $\hat m_a$ and $\hat m_b$
  of the magnetic moments of the ferromagnetic layers, like the
  standard equilibrium exchange interaction. On the other hand,
  the nonequilibrium
  torque mainly lies inside the plane spanned by $\vec m_a$ and
  $\vec m_b$.
\item
  For the Josephson-effect induced torque (or the standard
  equilibrium exchange interaction) to exist, transmission
  through the FNF junction needs to be phase coherent. Moreover,
  existence of the Josephson-effect induced torque requires that 
  majority and
  minority electrons experience different phase shifts upon
  transmission through or reflection from the ferromagnetic layers.
  The nonequilibrium torque, in contrast, only needs spin filtering 
  (different transmission or reflection probabilities for majority and
  minority electrons), while
  coherence is not important.\cite{wmbr,brataas}
\item
  Like the nonequilibrium torque, the Josephson-effect induced
  torque is carried by states close to the Fermi energy. The
  standard equilibrium torque has contributions from states 
  throughout the conduction band.
\item
  The equilibrium torques $\vec \tau^{\rm equ}_{a}$ and
  $\vec \tau^{\rm equ}_{b}$ on the moments of both ferromagnetic
  layers $F_a$ and $F_b$, respectivly, are equal in magnitude, 
  but opposite in direction, $\vec \tau^{\rm equ}_{a} = - \vec 
  \tau^{\rm equ}_{b}$. No
  such relation holds for the nonequilibrium torque.
\item 
  The sign and size of the nonequilibrium torque is controlled by the
  direction of the current. In contrast, the sign of the 
  Josephson effect induced torque is set by the
  superconducting phase difference $\phi$ and by the details of
  the scattering phase shifts from the ferromagnetic layers; it
  is not related to the direction of the supercurrent in any
  direct way. However, the order of magnitude of the Josephson 
  effect induced
  torque is set by the size of the critical supercurrent,
  $\tau^{\rm equ} \sim \hbar I_{\rm crit}/2e$.
\end{itemize}


We close with a discussion on the effect of the Josephson induced
exchange interaction on the dynamics of the magnetic moments. 
Typically, one of the two ferromagnetic moments (say the moment 
$\vec m_b$ of the layer $F_b$) is fixed by anisotropy forces, 
and the torque is
studied through its effect on the moment $\vec m_a$
of the ``free'' layer $F_a$.
The usual method to describe the dynamics $\vec m_a$ in the presence
of the current induced torque is via the phenomenological 
Landau-Lifshitz-Gilbert equation.\cite{slonc3,bazaliy}
The result of such a calculation is a critical value of the torque 
necessary for switching $\vec m_a$ with respect to the fixed moment
$\vec m_b$. This program was carried out for the 
non-equilibrium torque acting on a trilayer connected to two 
normal electrodes in Refs.\ \onlinecite{slonc3} and
\onlinecite{bazaliy}. 
The effect of the equilibrium torque considered here is simpler, as it admits 
a formulation in terms of the total energy of the system. Using
the phenomenological relation (\ref{tigre}), the 
Josephson induced exchange interaction corresponds
to an energy gain $\delta f$ per unit area equal to
\begin{equation}
\delta f=
\frac{\Delta_0}{\lambda_F^2}  \cos\phi \left[J_1 \hat m_a \cdot \hat m_b
+ J_2 (\hat m_a \cdot \hat m_b)^2\right].
\end{equation}
The criteria for switching the orientation of $\vec m_a$ is that the 
magnitude of this interaction energy exceeds that of the work done
against anisotropy forces acting on each
layer (arising from shape, crystalline structure, etc.). 
For a $2$ nm thick Cobalt layer, those
are of the order of $10^{-3}$ Jm$^{-2}$ and can be decreased by up to
two orders of magnitude for if Cobalt is replaced by Permalloy~\cite{textbook}
Ni$_{81}$Fe$_{19}$.
On the other hand, for $\Delta_0\approx 10K$ (as is the case for
Niobium), $\lambda_F\approx 1\AA$ and $J_1\approx 0.01$ (the value
found in our toy model simulations with slightly optimized values 
for the phase shift differences $\beta_a$ and $\beta_b$), the
Josephson induced interaction is of the order of $10^{-4}$ Jm$^{-2}$. 
Hence, we estimate that control of the relative orientation of 
$\vec m_a$ and
$\vec m_b$ should be experimentally accessible provided the local 
anisotropy forces are kept at a minimum. 

\begin{figure}[tbh]
\centerline{\psfig{figure=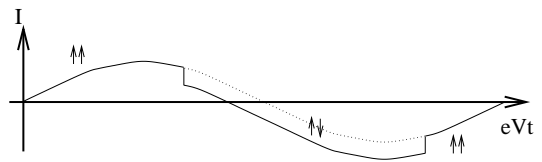,width=7.0cm}}
\vspace{3mm} \narrowtext
\caption[fig2]{
\label{otarie} Sketch of the current as a function of time when a small
voltage $V$ is applied across the junctions. When the Josephson effect
induced interaction exceeds the local anisotropy field, the relative
configurations of the two moments switches.}
\end{figure}

When the anisotropy forces are so small that switching of the
magnetic moments becomes a possibility, the ac Josephson effect
should provide a clear signature of the switching of the
ferromagnetic moments through the sensitivity of the supercurrent
to the angle $\theta$ between $\vec m_a$ and $\vec m_b$.
A possible scenario is sketched in Fig.~\ref{otarie}: the
equilibrium current observed in the Josephson effect 
should exhibit periodic shifts when the switchings occur. In order 
to evaluate the fastest time scale at which the switching of the
moments can occur, we return to the 
Landau-Lifshitz-Gilbert equation for the dynamics of the magnetic 
moments. As before, we suppose that $\vec m_b$ is kept fixed by 
a strong local anisotropy field, while the anisotropy field
acting
on $\vec m_a$ is negligible. Neglecting the bi-quadratic coupling
$J_2$, the 
dynamics of $\vec m_a$ can then be described by,\cite{slonc3,bazaliy}
\begin{equation}
\frac{\partial\hat m_a}{\partial t} =
\hat m_a \times \left[ \gamma H_J  
(\hat m_a\cdot\hat m_b) \hat m_b - \alpha \frac{\partial\hat
m_a}{\partial t}\right].
\end{equation}
Here $\gamma=g \mu_B/\hbar$ is the gyromagnetic ratio, $\alpha$ the
Gilbert damping coefficient and $H_J$ is the effective exchange
field representing the Josephson-effect induced torque,
\begin{equation}
H_J= \frac{J_1}{d_a |m_a|} \cos \phi (t),
\end{equation}
where $d_a$ is the
thickness of the layer $F_a$.
Neglecting the time dependance of the superconducting phase
difference $\phi$, this equation is easily 
solved,
\begin{equation}
\tan\theta=e^{-\Gamma t}\tan\theta_0 ,
\end{equation}
with $\Gamma= \gamma H_J \alpha/(1+\alpha^2)$. The switching
rate $\Gamma$ is maximum for $\alpha=1$ where, using the numerical 
values considered above and $|m_a|=1.6\ 10^6 {\rm A/m}$ (for cobalt), 
we find $\Gamma \approx 10$ GHz.
On the other hand, typical voltages used in ac Josephson
experiments are of the order of a few $\mu$V, which corresponds to 
frequencies of the order of a GHz. Therefore, for these frequencies, 
it should be possible, in principle, to observe the periodic 
switching of the
magnetization orientation as suggested in Fig.~\ref{otarie}.

\section*{ACKNOWLEDGMENT}

We thank  P. Chalsani, A.\ A.\ Clerk, M.\ Devoret, E.B.  Myers, D.\ C.\ Ralph and J.\ Voiron 
for  useful discussions. This work was supported by the
NSF under grant no.\ DMR 0086509 and by the Sloan foundation.



\appendix
\section*{Derivation of Eq.~(\ref{eq:dyson})}

\begin{figure}
\centerline{\psfig{figure=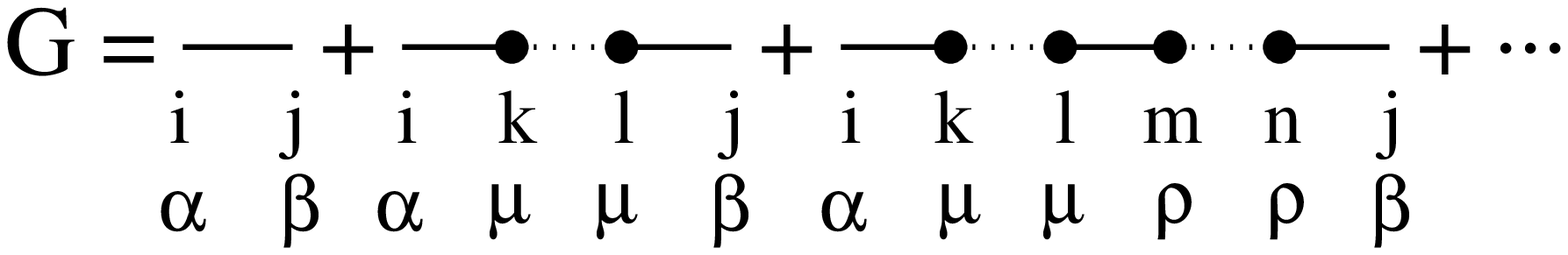,width=7.0cm}}
\vspace{-0.5cm}
(a)\bigskip

\centerline{\psfig{figure=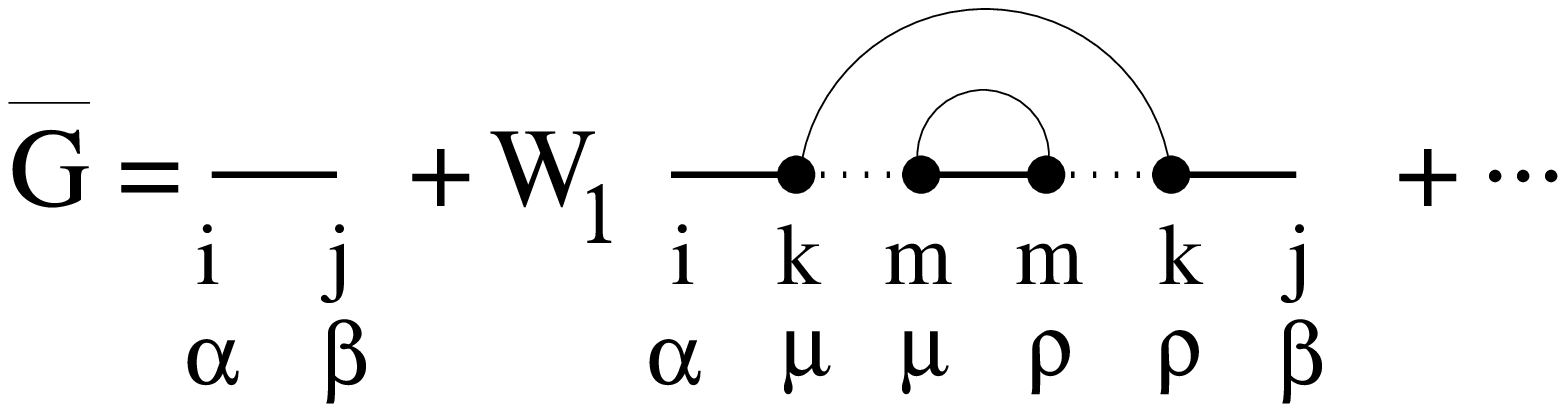,width=7.0cm}}
\vspace{-0.5cm}
(b)\bigskip

\centerline{\psfig{figure=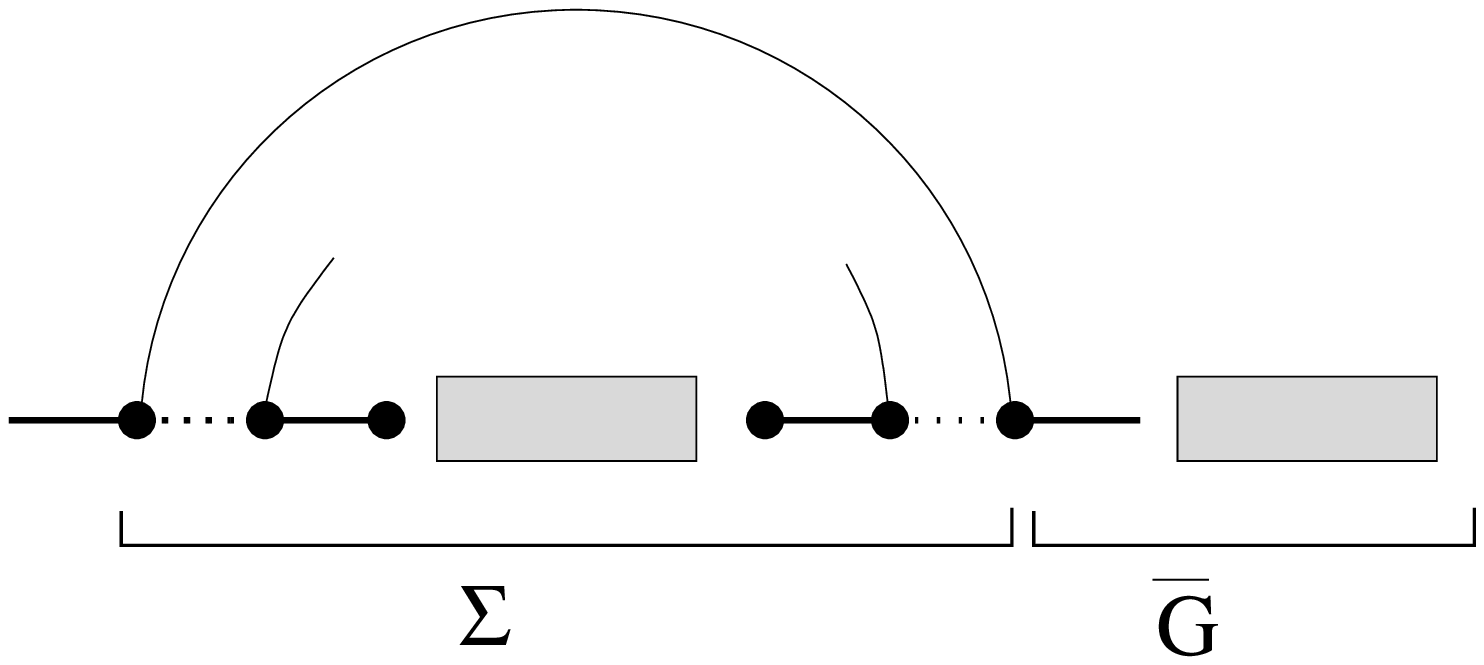,width=7.0cm}}
\vspace{-0.5cm}
(c)\bigskip
\hskip+1cm

\centerline{\psfig{figure=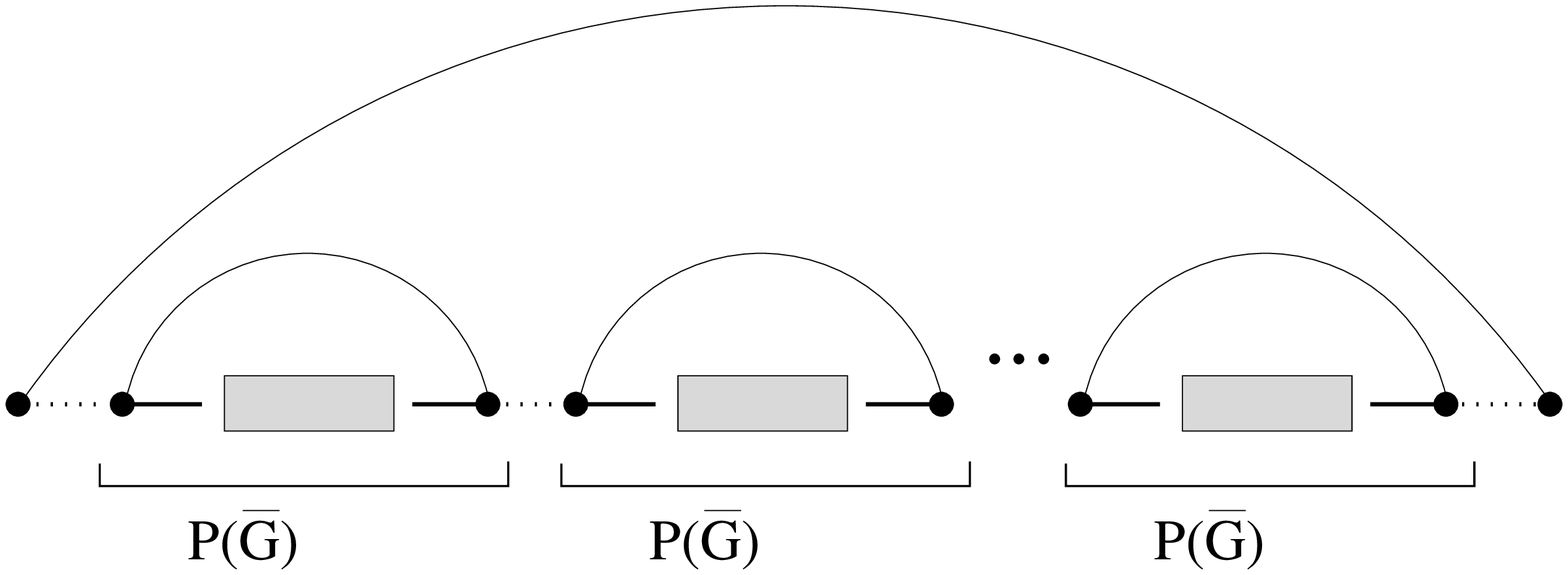,width=7.0cm}}
\vspace{-0.5cm}
(d)\bigskip

\caption{\label{fig:app} (a) Diagrammatic representation of Eq.\
(\protect\ref{eq:Gexpand}). (b) The first two diagrams
contribution to $\overline{\cal G}$. (c) General structure of
diagrams contributing to $\overline{\cal G}$. (d) General
structure of diagrams contributing to $\Sigma$.}
\end{figure}

In this appendix, we calculate the ensemble average $\overline {\cal
G}$ where ${\cal G}$ is defined in Eq.~(\ref{truite}). 
The ensemble average amounts to the calculation of an
average
over the circular orthogonal ensemble from random matrix theory
(the manifold of unitary symmetric matrices),
\begin{equation}
\overline{{\cal G}}= \int dS_0\  {\cal S}_{FS} 
  \frac{1}{1-{\cal S}_N {\cal S}_{FS}},
\end{equation}
where, in particle/hole ($e-h$) grading, 
${\cal S}_{N} = \mbox{diag}\,(S_0 \otimes 
\openone_2,S_0^* \otimes \openone_2)$, $\openone_2$ being the 
$2 \times 2$ identity matrix in spin space and $S_0$ being an 
$2N_{\rm ch} \times 2N_{\rm ch}$
symmetric unitary matrix, and $dS_0$ is the invariant measure for
integration over the circular orthogonal ensemble. 
To perform the average over $S_0$, we
use the diagrammatic technique of Ref.~\onlinecite{brouwer1} and
calculate $\overline{{\cal G}}$ to leading order in $1/N_{\rm ch}$.
First, ${\cal G}$ is expanded in powers of ${\cal S}_{FS}$, 
\begin{equation}
{\cal G}={\cal S}_{FS} + {\cal S}_{FS} {\cal S}_N{\cal S}_{FS} +
  {\cal S}_{FS} {\cal S}_N{\cal S}_{FS} {\cal S}_N{\cal S}_{FS}
  +\ldots
  \label{eq:Gexpand}
\end{equation}
and the corresponding terms are associated with diagrams: a full
line corresponds to a factor ${\cal S}_{FS} $ and a dotted line to 
factor ${\cal S}_N$, see Fig.\ \ref{fig:app}a. At each dot one
sums over a latin index ranging from $1$ to $2 N_{\rm ch}$, 
representing the channel space, and a greek index ranging
from $1$ to $4$, representing the combination of $e-h$ and spin space.
According to the diagrammatic rules of Ref.\ \onlinecite{brouwer1},
the average is then done by connecting all dots by thin
lines, summing over all possible ways of pairing up the dots. 
To leading order in $1/N_{\rm ch}$, only planar diagrams 
contribute, 
i.e., the diagrams where the thin lines do not cross. When two dots
are connected, the corresponding latin indices are identified and
summed over, and the
constraint is imposed that two greek indices involved
have to be different 
in $e-h$ space (i.e., if one index corresponds to $e$, the other one 
has to correspond to $h$). 
The rationale for this constraint is that only
contractions that involve $S_0$ and its complex conjugate are allowed
in the average.
Finally, a weight factor is associated with each diagram: each
cycle formed by an alternation of dotted lines and thin solid lines in 
the diagram contributes a factor $W_i$, where $i$ is equal to half the 
number of
dotted lines contained in the cycle. The $W_i$ are tabulated in 
Ref.~\onlinecite{brouwer1}; for the purpose of this integral,
we only need their generating function\cite{brouwer2}
\begin{equation} 
\sum_{i=1}^{\infty} W_i z^{i-1}= \left(\sqrt{(2 N_{\rm ch})^2 +4z} -(2N_{\rm
ch})\right)/2z. \label{eq:generating}
\end{equation}
The weight factors $W_i$ with $i > 1$ correspond to contributions
to the average that go beyond a Gaussian evaluation using Wick's
theorem.

The first non vanishing diagrams are shown in Fig.\ \ref{fig:app}b.
They correspond to
\begin{equation}
\overline{G}={\cal S}_{FS} + 2 N_{\rm ch} W_1 
\  {\cal S}_{FS}  \ {\bf P}( {\cal S}_{FS})
\ {\cal S}_{FS}  +\ldots \label{eq:app2}.
\end{equation}
The projector operator ${\bf P}$ was introduced in
Eq.~(\ref{perroquet}). It implements the constraint that only
greek indices representing $e$ and $h$ degrees of freedom
can be contracted.

Now, we are ready to obtain the self-consistent equation
(\ref{eq:dyson}) for
$\overline{G}$. Except from the zeroth order (first term in
Eq.\ (\ref{eq:app2}), all the diagrams
involved in $\overline{G}$ have the structure shown in 
Fig.\ \ref{fig:app}c,
where the boxes stand for all possible allowed contractions. 
The
important point here is that 
there is no thin line connecting the two different boxes 
since we are only considering  planar diagrams. 
The sum over
all the different contractions represented by the right box 
gives all the possible diagrams and therefore equals
$\overline{{\cal G}}$ itself. Denoting the left part of
Fig.~\ref{fig:app}c 
by $\Sigma$, we have
\begin{equation}
\overline{{\cal G}}= {\cal S}_{FS}  + {\cal S}_{FS}  
\Sigma \overline{{\cal G}}, 
\end{equation}
which is Eq.\ (\ref{chimpanze}).
The diagrams contributing to $\Sigma$ have the structure shown in
Fig.\ \ref{fig:app}d. They contain $2n-1$ building blocks,
$n=1,2,\ldots$, with weight factor $W_n$. 
(Only odd numbers appear, because
of the constraint that only indices belonging to $S_0$ and its 
complex conjugate can be contracted.) Each building block can be
identified with ${\bf P}({\cal \overline{G}})$. Hence,
\begin{equation}
\Sigma = \sum_{n=1}^{\infty} W_n {\rm Tr} 
\ \left[ 2 N_{\rm ch} {\bf P}(\overline{{\cal G}})\right]^{2n-1},
\end{equation}
which leads to Eq.~(\ref{ouistiti}) if we use the generating function
(\ref{eq:generating})
for the weight factors $W_n$.



\begin{thebibliography}{99}
\bibitem{propoverbeux} A Co-Cu-Co trilayer might not be the best choice
of materials for this effect because the Curie temperature of Co is
rather large. However, to our knowledge, no ab-initio
data for weakly ferromagnetic trilayers  are available in the literature.
\bibitem{brouwer2} The generating function can be easily obtained by
calculating the average of $(z - S_0 S_0^{\dagger})^{-1}$.
\end{thebibliography}
\end{document}